\documentstyle[12pt,twoside]{article}
\input amssym.def
\input amssym

\def\R{{\Bbb R}}
\def\N{{\Bbb N}}
\def\eps{\varepsilon}
\def\skippar{\par\smallskip}
\def\phi{\varphi}

\newtheorem{theorem}{Theorem}
\newtheorem{lemma}{Lemma}

\begin{document}

\title{Escape Orbits for Non-Compact Flat Billiards}
\author{
	Marco Lenci\thanks{On leave from: Dipartimento di Matematica, 
	Universit\`a di Bologna, 40127 Bologna, Italy} \\
	Mathematics Department \\
	Princeton University \\
	Princeton, NJ \ 08544 \\
	U.S.A. \\
	{\tt\small marco@math.princeton.edu}
}
\date{February 1996}	

\maketitle

\begin{abstract}
	It is proven that, under some conditions on $f$, the non-compact flat 
	billiard $\Omega = \{ (x,y) \in \R_0^{+} \times \R_0^{+};\ 0\le y \le 
	f(x) \}$ has no orbits going {\em directly} to $+\infty$. The 
	relevance of such sufficient conditions is discussed.
\end{abstract}

\section{Introduction}

Let $f$ be a smooth function from $\R_0^{+}$ to $\R^{+}$, bounded, 
vanishing when $x\to +\infty$. No integrability assumptions are given. 
Now construct a plane billiard table in the following way: 
$\Omega = \{ (x,y) \in \R_0^{+} \times \R_0^{+};\ 0\le y \le f(x) \}$ 
as is shown in figure 1. Imagine to have a dimensionless 
particle moving freely within $\Omega$ and reflecting elastically 
at its boundary.  
\par 
Some interest seems to be focused recently on these (see 
\cite{l,ss,k,ddl}) and other billiard systems over non-compact 
manifolds (intriguing and related examples include \cite{b,gu,ga}).  
In this contest an issue regards whether and how a trajectory may 
leave any compact region of the configuration space. The study of 
this is of theoretical interest in itself, as well serving as a possible 
starting point for the investigation of the ergodic properties of 
non-compact dynamical systems of which not much seems to be known.  
Plus, and this is the main motivation for recent work, it has a direct 
link with the search for {\em quantum chaos} in the quantized 
version of these systems.  Specifically,
unbounded orbits represent the other side of the coin compared to 
periodic orbits. The distribution of the latter provides 
information on the spectrum of the Laplace-Beltrami (i.e.  
Hamiltonian) operator on the given manifold via the celebrated 
Gutzwiller trace formula or similar asymptotic expansions \cite{g,st}.  
Being able to ``count'' periodic orbits means gaining information on 
the aysmptotics of the quantum energy levels. This has suggested the 
idea of {\em coding} periodic orbits, i.e. associating an infinite 
string of symbols to each periodic orbit, possibly in a bijective 
way.  
\footnote{There are quite a few works on that. \cite{gu} has good 
analogies with the present paper.}
Understandably, unbounded trajectories represent a nuisance 
for such a coding. Finally, it is known that the heat kernel 
expansion is sensitive to finite or infinite cusps \cite{st}. One 
would like to relate this fact to the presence of orbits eventually 
falling into the cusp.
\par
The question we ask ourselves here is the following: are there any orbits 
of our billiard in $\Omega$ which go {\em directly} to infinity, i.e.  
maintaining for all times a positive $x$-component of their velocity?  
To fix the notation let us call such trajectories {\em escape orbits}.  
Of course there is always one of these: it is the orbit which lies on 
the horizontal semi-axis. We call it the {\em trivial orbit}.  
\par 
We are going to show that no other orbits can share this property, 
provided we require some conditions on $f$. The question was first
touched on by A.M.Leontovich  
\footnote{To my knowledge.}
in 1962 (\cite{l}, Theorem 2) where -- even though he was searching 
for oscillating unbounded orbits -- he obtained the above result for 
{\em eventually convex} $f$'s, i.e. convex in a neighborhood of 
$\infty$. This is the same result we find much more recently 
in J.L.King's review \cite{k}. It can be explained easily, at least 
for billiard tables of finite area. In that case the cusp has 
asymptotically a vanishing measure. Now, if we have an escape orbit, 
then, due to the hyperbolicity of the flow in that region of the 
phase space, we can find a non-zero measure set of escape orbits, 
that is which go into the cusp, with an obvious contradiction.
\footnote{This argument is described in \cite{k} as ``A gallon of water 
won't fit inside a pint-sized cusp''.} 
In fact we can always fix, as the initial point of our escape orbit,  
a point $(x_0,y_0) = (x_0,f(x_0))$ on the upper boundary of $\Omega$, 
far enough to lie in the region where $f$ in convex. The initial velocity 
will have an $x$-component $v_x > 0$. Now it is easily seen that every 
other set of initial conditions $x'_0 \ge x_0$ and $v'_x \ge v_x$ (provided 
$v$ and $v'$ have the same modulus) would lead to a new escape orbit, due 
to the dispersing effect of the convex upper wall. 
The same argument may be used to deduce that an infinite cusp on the 
Poincar\'e's disc does not allow non-trivial trajectories to collapse 
into it, which is implicitly stated in \cite{gu}.
\par
We are going to relax the hypothesis for the non-existence result to 
hold: asymptotic hyperbolicity is not a necessary condition at all. 
We may allow $f$ to have flex points and abrupt negative variations 
(see, for instance, figure 2c). The proofs are presented in the
next section, while examples of non-convex $f$'s fulfilling our 
hypotheses are discussed in the last section in order to understand what 
the new assumptions actually mean and how far they can be pushed.

\section{The result}

\begin{theorem}
	Consider the plane billiard table generated as above by the function
	$f$ defined on $\R_0^{+})$, twice differentiable, positive, bounded, 
	such that
	$$
		f(x) \searrow 0 \mbox{ as } x\to +\infty.
		\eqno{{\em (H1)}}
	$$
	Also, for $x$ sufficiently large,
	$$
		f'(x) < 0,
		\eqno{{\em (H2)}}
	$$
	Then, under either one of these assumptions:
	$$
		\limsup \frac{f'(x)}{f(x)} <0;
		\eqno{{\em (A1)}}
	$$
	or
	$$
		\lim_{x\to +\infty} f'(x) =0 \mbox{ and } \limsup_{x\to +\infty} 
		\frac{f'' f}{f'}(x) < +\infty;
		\eqno{{\em (A2)}}
	$$
	no orbits but the trivial one are escape orbits.
\end{theorem}
\skippar
{\sc Remark.} The assumption about the convexity of $f$ is contained in 
(A2): in fact if $f'' \geq 0$, then necessarily $f' \nearrow 0$ and 
$f'' f/f' \leq 0$.
\skippar
{\sc Proof.} Suppose, contrary to our goal, we have a non-trivial escape 
orbit: let us fix without loss of generality an initial point in a 
neighborhood of $+\infty$ where all the asymptotic hypotheses hold. 
For instance, (H2) would do, and (A2), if this is the case, would be 
read as
\begin{equation}
	f'(x) \geq -\eps \mbox{ and } \frac{f'' f}{f'}(x) < k_1;
	\label{newA2}
\end{equation}
for some $\eps>0$. Also, for some consistency of notation let us 
suppose the initial point is a bouncing point on the upper wall, i.e. 
$(x_0,y_0) = (x_0,f(x_0))$.
\skippar
Using the notations of figure 1 we have the fundamental relation:
\begin{equation}
	\tan\theta_{n+1} (x_{n+1}-x_n) = f(x_n) + f(x_{n+1}).
	\label{fundamental}
\end{equation}
With a bit of geometry, looking at the same picture, we get
\begin{displaymath}
	\theta_{n+1} = \theta_n + \pi - 2\alpha_n = \theta_n + 2\delta_n,
\end{displaymath}
calling $\delta_n = - \arctan(f'(x_n)) > 0$. This summarizes to
\begin{displaymath}
	\theta_n = \theta_1 + 2\ \sum_{k=1}^{n-1} \delta_k.
\end{displaymath}
Thus $\{ \theta_n \}_{n \ge 1}$ is an increasing sequence. Since we 
have assumed the particle never go backwards, then $\theta_n < 
\pi/2$ for all $n \ge 1$, so $\theta_n \nearrow 
\theta_\infty \in [\theta_1,\pi/2]$. Hence $\tan\theta_n \ge 
\tan\theta_1 =: k_2 > 0$. From this, the monotonicity on $f$, and 
(\ref{fundamental}) we have
\begin{equation}
	x_{n+1} - x_n \leq \frac{2}{k_2} f(x_n).
	\label{estimate1}
\end{equation}
What stated above implies that $\sum_k \delta_k < +\infty$. Therefore 
$\delta_k \to 0$. As a consequence, we see that $\exists k_3\in ]0,1[$ 
such that $\delta_k = \geq k_3 \, \tan\delta_k = k_3 |f'(x_k)|$. 
If we are able to prove that 
\begin{equation}
	-\sum_{k=0}^{\infty} f'(x_k) = +\infty,
\label{to-violate}
\end{equation}
that inequality implies that $\sum_k \delta_k$ cannot converge, 
creating a contradiction which finishes the proof.
\skippar
Defining $g:= -f'/f >0$ will greatly simplify our notation.
From (\ref{estimate1}), we obtain, for some constant $k_4$,
\begin{equation}
	-\sum_n f'(x_n) \geq k_4 \sum_n g(x_n) (x_{n+1} - x_n).
	\label{estimate2}
\end{equation}
\skippar
{\sc Case} (A1). Obviously (A1) reads $g \geq k_5 > 0$. Hence, since 
by hypothesis $x_n \to\infty$, (\ref{estimate2}) gives 
(\ref{to-violate}). It may worth remind that (A1) means we have 
exponential decay for $f$. In fact, after an integration, we see 
that $\forall x_2 > x_1 \geq 0$,
\begin{displaymath}
	f(x_2) \leq f(x_1)\: e^{-k_5(x_2 - x_1)}.
\end{displaymath}
\skippar
{\sc Case} (A2). The proof is little more involved here. We use our 
assumption on the limit of $f'$ to prove the following
\begin{lemma}
	There exists a constant $k_6$ such that $\forall n\in\N$
	\begin{displaymath}
		\frac{f(x_n)}{f(x_{n+1})} \leq k_6.
	\end{displaymath}
\end{lemma}
{\sc Proof.} Let us call $\bar x_n$ the point in $[x_n,x_{n+1}]$ provided
by the Lagrange mean value theorem. We can write
\begin{displaymath}
	\frac{f(x_n)}{f(x_{n+1})} = 1 - \frac{f'(\bar x_n) (x_{n+1}-x_n)}
	{f(x_{n+1})}.
\end{displaymath}
Using (\ref{estimate1}), this is turned into
\begin{displaymath}
	\frac{f(x_n)}{f(x_{n+1})} \left( 1 + \frac{2 f'(\bar x_n)}{k_2} \right) 
	\leq 1,
\end{displaymath}
which yields the lemma, since the term in the brackets is positive 
for $n$ large enough, because of the assumption about the vanishing of 
$f'$.
\skippar
This enables us to prove
\begin{lemma}
	There exists a constant $k_7 \in ]0,1[$ such that, for $n$ 
	sufficiently large, $g(x_n) \geq k_7 \max_{[x_n,x_{n+1}]} g$.
\end{lemma}
{\sc Proof.} Proving the statement is equivalent to proving that we can 
find a $k_8 > 0$ for which
\begin{displaymath}
	\log g(\tilde x_n) - \log g(x_n) \leq k_8,
\end{displaymath}
where $\tilde x_n$ maximizes $g$ in $[x_n,x_{n+1}]$. Using again the 
Lagrange mean value theorem, the fact that $\tilde x_n-x_n \le (2/k_2)
f(x_n)$ (a consequence of (\ref{estimate1})), and the previous lemma, we 
obtain
\begin{eqnarray*}
	\log g(\tilde x_n) - \log g(x_n) & = & \frac{g'}{g}(\hat x_n) 
	(\tilde x_n-x_n) \\
	& \leq & \frac{2}{k_2} \left( \frac{f''}{f'} - \frac{f'}{f} 
	\right) (\hat x_n)\ f(x_n)  \\
	& \leq & k_9 \left( \frac{f''}{f'} - \frac{f'}{f} \right) (\hat x_n) 
	\ f(x_{n+1})  \\
	& \leq & k_9 \left( \frac{f''f}{f'} - f' \right) (\hat x_n) \leq k_9 
	(k_1 + \eps),
\end{eqnarray*}
having used (\ref{newA2}) in the last step.
\skippar
We are now prompted to get (\ref{to-violate}) in this case, too. 
Looking at (\ref{estimate2}) we can write:
\begin{eqnarray*}
	\sum_n g(x_n) (x_{n+1} - x_n) & \ge & k_7 \sum_n
	(\max_{[x_n,x_{n+1}]} g) (x_{n+1} - x_n)  \\
	& \ge & k_7 \int_{x_0}^{+\infty} g(x)dx = +\infty,
\end{eqnarray*}
since $-\int^\infty (f'/f) = -\lim_{x\to\infty} (\log f(x) + const) =
+\infty$. This ends the proof of the theorem.

\section{Discussion}

The obvious news the theorem says, compared to the mentioned condition 
$f''>0$, is the possibility for $f'$ to oscillate, 
to a certain extent. Dynamically speaking, the change in direction 
our particle gets every time it bounces against the upper wall 
($\delta_k = -\arctan(f'(x_k))\,$) need not be a monotone sequence. As 
a matter of fact, (A2) precisely controls the amount of such an
oscillation. An example will illustrate the case: for $\alpha >1, 
\beta >0, c > 1$ define $f'_1(x) := -x^{-\alpha} (\sin(x^{\beta}) + c) 
< 0$.  This means that we define $f_1(x) := -\int_{x}^{\infty} f'_1 
(z)dz$, which makes sense as a convergent integral. Therefore 
$f''_1(x) = -\beta x^{-\alpha+\beta-1} \cos(x^{\beta}) + 
O(x^{-\alpha-1})$, showing that $f_1$ is not convex. Now, the 
asymptotic behavior of $f_1$ and its derivatives is easily extracted 
to yield
\begin{displaymath}
	\frac{f''_1 f_1}{f'_1}(x) \asymp x^{-\alpha+\beta}.
\end{displaymath}
Thus, (A2) holds if, and only if, $\alpha \ge \beta$, meaning that the 
faster $f_1$ vanishes the more violent the oscillation of $f'_1$ is 
allowed to be.
\par
Another example may be interesting to present, to show that there are 
cases where (A1) holds but (A2) does not. Pick up a $\phi \in 
C^{\infty}(\R)$ supported in $]-1/2,1/2[$ with $\int \phi = 1-e^{-1}$.  
Call, for $k \in\N$,
\begin{displaymath}
	\phi_k(x) := \phi((x-k-1/2)\, e^k),
\end{displaymath}
supported in $]k+1/2-e^{-k}/2,k+1/2+e^{-k}/2[$. Let us now define 
$h(x) := \sum_{k=0}^{\infty} \phi_k(x)$.  The result is shown in 
figure 2a. We see that
\begin{displaymath}
	\int_k^{k+1} h(x)dx = \int_k^{k+1} \phi_k(x)dx = (1-e^{-1}) e^{-k}.
\end{displaymath}
Also denote by $H(x) := \int_{x}^{\infty} h(z)dz$. Finally, let us 
introduce $f'_2(x) := -e^{-x} -h(x)$, corresponding to $f_2(x) = 
e^{-x} + H(x)$. Their graphs are displayed in figure 2b and 2c, 
respectively. Certainly $f'_2 \not\to 0$ and (A2) is not verified. 
Now, from above we can estimate the value of $f_2$. In fact 
$e^{-[x]-1} \leq H(x) \leq e^{-[x]}$ giving $H(x) \leq e^{-x+1}$. 
Therefore
\begin{displaymath}
	\frac{|f'_2|}{f_2}(x) = \frac{e^{-x}+h(x)}{e^{-x}+H(x)} \geq 
	\frac{e^{-x}}{e^{-x}+e^{-x+1}} = \frac{1}{1+e} \geq 0.
\end{displaymath}
That is: (A1) holds as well as the result, in this case.
\skippar
Is it difficult to say to what extent our theorem is inclusive of the 
general case or how it can be refined. The point here is that finding 
a sufficient condition for the non-existence 
of an escape orbit is much more direct than 
finding a necessary condition. The shape of $f$ can be {\em pathological} 
enough but not in a suitable way that allows a trajectory to go directly
to infinity. One thing can be said, though: hypotheses (H1) and (H2) do 
not suffice and one needs some extra assumption to control a possible 
wild behaviour of $f'$. As a matter of fact we may now sketch the 
construction of a billiard table verifying those hypotheses and having one 
escape orbit. We will start by first drawing the orbit and then a compatible 
$f$. 
\par
Consider the polyline shown in figure 3a with $\theta_1 \in ]0,\pi/2[$ and
$\{y_n \}$ any non-integrable sequence such that $y_n \searrow 0$. If this 
were an escape orbit then we would have $f(x_n) = y_n,\, f'(x_n) = 0$ and 
$\theta_n = \theta_1\ \forall n$. Furthermore $x_{n+1} - x_n = 
(y_{n+1} + y_n)/\tan\theta_1$ so that $\lim_{n\to\infty} x_n = \infty$, 
because of our assumption on $\{y_n \}$. Of course any $f$ giving rise
to such an orbit cannot satisfy (H2), because of the flat tangent at 
the bouncing points, but we can slightly modify 
our picture in order to fit it. Take an 
integrable sequence $\{\delta_n \},\: \delta_n>0$ such that 
$\theta_\infty := \theta_1 + 2 \sum_n \delta_n < \pi/2$. 
Now modify the trajectory in figure 3a, ``shrinking''
it in order to have $\theta_n := \theta_1 + 2 \sum_{k=1}^{n-1} \delta_k$;
keep $y_n$ fixed. The result is drawn in figure 3b. This is again an
escape orbit since, due to our choice of $\theta_\infty$, the contraction
of the little triangles has a lower bound, i.e. $x_{n+1} - x_n \geq
(y_{n+1} + y_n)/\tan\theta_\infty$. One can now very easily construct 
an $f$ which satisfies (H1) and (H2) and whose graph is an upper wall 
for this trajectory.
\par
This proves our remark. 

\section{Acknowledgments}

I wish to thank Ya.G.Sinai, C.Liverani, E.Gutkin and N.Chernov for 
their kindness in discussing with me about this subject. I also 
express my gratitude to A.Parmeggiani for his support, during 
the preparation of this paper.


\begin{thebibliography}{9}
	\bibitem{l}  {\sc A.M.Leontovich}
	\newblock {\it The Existence of Unbounded Oscillating Trajectories in 
	a Problem of Billiards}
	\newblock Sov. Acad. Sci., doklady. Math. {\bf 3},4 (1962) 1049-1052

	\bibitem{g}  {\sc M.C.Gutzwiller}
	\newblock {\it Chaos in Classical and Quantum Mechanics}
	\newblock Springer-Verlag, New York, 1990

	\bibitem{b}  {\sc P.M.Bleher}
	\newblock {\it Statistical Properties of a Particle Moving in a 
	Periodic Scattering Billiard}
	\newblock in: Stochastic Methods in Experimental Science, W.Kasprzak 
	A.Weron eds., World Sci. Publ., River Edge, NJ, 1990
	
	\bibitem{gu}  {\sc M.J.Giannoni, D.Ullmo}
	\newblock {\it Coding Chaotic Billiards. I. Non Compact Billiards on 
	a Negative Curvature Manifold}
	\newblock Physica D {\bf 41} (1990) 371-390

	\bibitem{ss}  {\sc M.Sieber, F.Steiner}
	\newblock {\it Classical and Quantum Mechanics of a Strongly Chaotic 
	Billiard System}
	\newblock Physica D {\bf 44} (1990) 248-266 
	
	\bibitem{st}  {\sc F.Steiner, P.Trillenberg}
	\newblock {\it Refined Asymptotic Expansion for the Partition 
	Function of Unbounded Quantum Billiards}
	\newblock J. Math. Phys. {\bf 31},7 (1990) 1670-1676
	
	\bibitem{ga}  {\sc G.A.Galperin}
	\newblock {\it Asymptotic Behavior of a Particle in a Lorenz Gas}
	\newblock Russ. Math. Surveys {\bf 47},1 (1992) 258-259
	
	\bibitem{k}  {\sc J.L.King}
	\newblock {\it Billiards Inside a Cusp}
	\newblock Math.Intell. {\bf 17},1 (1995) 8-16
	
	\bibitem{ddl}  {\sc M.Degli Esposti, G.Del Magno, M.Lenci}
	\newblock {\it Escape Orbits for a Class of Infinite Step Billiards}
	\newblock In preparation
\end{thebibliography}
\end{document}